\pdfoutput=1

\documentclass[aps,preprintnumbers,amsmath,amssymb,twocolumn, tightenlines,superscriptaddress]{revtex4}

\usepackage{graphicx}
\usepackage{epsfig}

\newcommand{\be}{\begin{eqnarray}}
\newcommand{\ee}{\end{eqnarray}}

 \newcommand{\gsim}{\mathrel{\hbox{\rlap{\lower.55ex \hbox {$\sim$}}
                   \kern-.3em \raise.4ex \hbox{$>$}}}}
\newcommand{\lsim}{\mathrel{\hbox{\rlap{\lower.55ex \hbox {$\sim$}}
                   \kern-.3em \raise.4ex \hbox{$<$}}}}

\newcommand{\ba}{\begin{eqnarray}}
\newcommand{\ea}{\end{eqnarray}}

\begin{document}


\title{Pairing Phase Transitions of Matter under Rotation}
\author{Yin Jiang} 
\address{Physics Department and Center for Exploration of Energy and Matter,
Indiana University, 2401 N Milo B. Sampson Lane, Bloomington, IN 47408, USA.}
\author {Jinfeng Liao} 
\address{Physics Department and Center for Exploration of Energy and Matter,
Indiana University, 2401 N Milo B. Sampson Lane, Bloomington, IN 47408, USA.}
\address{RIKEN BNL Research Center, Bldg. 510A, Brookhaven National Laboratory, Upton, NY 11973, USA.}
\date{\today}

\begin{abstract}
The phases and properties of matter under global rotation have attracted much interest recently. In this paper we investigate the pairing phenomena in a system of fermions under the presence of rotation. We find that there is a generic suppression effect on pairing states with zero angular momentum. We demonstrate this effect with  the chiral condensation and the color superconductivity in hot dense QCD matter as explicit examples.  In the case of chiral condensation, a  new phase diagram in the temperature-rotation parameter space is found, with a nontrivial critical point.
\end{abstract}

\maketitle

{\it Introduction.---}The phases and properties of matter can become  highly nontrivial under  rotation, and have attracted a lot of interest recently. Such studies bear particular relevance for the strongly interacting matter of Quantum Chromodynamics (QCD). For example,  astrophysical objects like neutron stars, made of dense QCD matter, can be rapidly spinning~\cite{Watts:2016uzu,Grenier:2015pya}. In relativistic heavy ion collision experiments, the typical collision events are off-central and the created QCD matter will carry a nonzero angular momentum~\cite{Liang:2004ph,Becattini:2007sr,Csernai:2013bqa,Jiang:2016woz,Deng:2016gyh}. There has also been impressive progress to study the rotating QCD matter using lattice gauge theory simulations~\cite{Yamamoto:2013zwa}.

It is found that in rotating matter, many interesting transport phenomena could occur. For example, the fluid rotation (as quantified by a nonzero vorticity) can induce certain anomalous transport processes in a system of chiral fermions, with the notable examples of chiral vortical effect~\cite{Kharzeev:2007tn,Son:2009tf,Kharzeev:2010gr} and chiral vortical wave~\cite{Jiang:2015cva}. These can lead to measurable experimental signals (see e.g. recent reviews in \cite{Kharzeev:2015znc,Liao:2014ava}).   In the study of such anomalous transport, it has been identified that the fluid rotation plays a very analogous role to an external magnetic field. Indeed there appears to be an interesting analogy between the chiral vortical effect and the so-called chiral magnetic effect~\cite{Kharzeev:2007tn,Kharzeev:2007jp}, as well as between the chiral vortical wave and the so-called chiral magnetic wave~\cite{Kharzeev:2010gd,Burnier:2011bf}.

Apart from transport properties, it is of significant interest to explore the effects of rotation on the phase structures and phase transitions of matter in both relativistic and non-relativistic cases. In particular, it is known that an external magnetic field can bring interesting effects on the thermodynamics and phase diagram on e.g. QCD matter~\cite{Bali:2011qj,Bali:2012zg,Fukushima:2012kc,Kojo:2012js,Chao:2013qpa,Cohen:2013zja}, with the well-known example of magnetic catalysis and inverse catalysis (see reviews in e.g. \cite{Shovkovy:2012zn,Miransky:2015ava}) on the chiral condensation. Given the close analogy between rotation and magnetic field, it is tempting to ask whether and how the rotation could influence the various phase transitions. In this paper,  we investigate the pairing phenomena in a system of fermions under the presence of rotation. We will show that there is a generic suppression effect on pairing states with zero angular momentum. We demonstrate this effect with  the chiral condensation and the color superconductivity in hot dense QCD matter as explicit examples.

{\it Rotational Suppression Effect on Scalar Pairing States.---} Before going to more detailed computation, let us first explain, in an intuitive way, the generic rotational suppression effect on scalar pairing states. We are considering in general a system of spin-$\frac{1}{2}$ fermions. They could be e.g. the dense quark or nucleon matter in the context of compact stars~\cite{Chen:2015hfc, Berti, Demorest} or the cold atomic gases~\cite{Fetter:2009zz,2008PhRvA..78a1601U,2009PhRvA..79e3621I}. More conventional examples include e.g. electrons or holes in solid state systems, liquid helium-3, etc.  The pairing phenomenon between fermions under suitable conditions encompasses a wide range of systems. Examples include e.g. electron-electron pairing in superconductors, atom-atom pairing in helium-3 or cold fermi gases, nucleon-nucleon pairing in large nuclei or dense nuclear matter, quark-anti-quark pairing in the chiral condensate of QCD, or quark-quark pairing in color superconductivity, etc. We focus on {\em the scalar pairing states}, i.e. states which have zero total angular momentum. Note that for a pair of spin-$\frac{1}{2}$ fermions, there are different ways of forming a spin-0 pairing state: either, the pair could have both nonzero orbital angular momentum $L$ and nonzero total spin $S$, with $L$ and $S$ being opposite thus resulting in total $J=0$; or the pair could have zero orbital angular momentum, and have opposite individual spin configurations for the two fermions.

As we will show below, when such a system is under rotation, there will be a generic rotational suppression effect on the scalar pairing states. Intuitively this can be understood as follows. The global rotation, implying a nonzero macroscopic angular momentum of the whole system,  will induce a {\em rotational polarization effect} which tends to ``force'' all microscopic angular momentum to be   aligned with the global angular momentum. So for a pair of fermions, their relative orbital angular momentum $L$ as well as their individual spins would prefer to be parallel to the global angular momentum rather than to arrange themselves into a scalar state with zero angular momentum. This therefore leads to a generic suppression effect on the scalar pairing states. It also implies that pairing states with nonzero angular momentum could become more favorable. In the following, we quantitatively demonstrate this effect with two nontrivial examples in the QCD matter: the chiral condensate (of quark-anti-quark pairing states with $L=S=1$ but $J=0$) and the diquark condensate (of quark-quark pairing state with $L=S=0$).

{\it Description in Rotating Frame.---} Let us consider a system of spinor particles that is under very slow rotation with a constant angular velocity denoted by $\vec \omega$ along a certain fixed axis. This system can be equivalently described as a system at rest in a {\em rotating reference frame}, see e.g. discussions in e.g.~\cite{Yamamoto:2013zwa,Fetter:2009zz}. We denote space-time as $(t,\vec x)$ with flat Minkowski metric $\eta_{\mu\nu}=Diag(1,-1,-1,-1)$. The local velocity of this rotating frame (with respect to the original non-rotating frame) is given by $\vec v = \vec \omega \times \vec x$. The space-time metric of the rotating frame becomes a curved one, given by:
\begin{eqnarray}
g_{\mu\nu}=\left(
\begin{array}{cccc}
 1-{\vec v}^{\, 2} & -v_1 & -v_2 & -v_3 \\
 -v_1 & -1 & 0 & 0 \\
 -v_2 & 0 & -1 & 0 \\
 -v_3 & 0 & 0 & -1 \\
\end{array}
\right)
\end{eqnarray}
In such description, the usual (free) Dirac Lagrangian for spinor gets modified to take the following form:
\begin{eqnarray}
{\cal L} =  \bar{\psi} \left[ i \bar{\gamma}^\mu  (\partial_\mu+ \Gamma_\mu ) - m \right] \psi
\end{eqnarray}
where $m$ is the fermion mass. The $\bar{\gamma}^\mu=e_{a}^{\ \mu} \gamma^a$ with $e_{a}^{\ \mu}$ the tetrads for spinors and $\gamma^{a}$ the usual Dirac $\gamma$ matrices. The spinor connection is given by $\Gamma_\mu=\frac{1}{4}\times\frac{1}{2}[\gamma^a,\gamma^b] \, \Gamma_{ab\mu}$ where $\Gamma_{ab\mu}=\eta_{ac}(e^c_{\ \sigma} G^\sigma_{\ \mu\nu}e_b^{\ \nu}-e_b^{\ \nu}\partial_\mu e^c_{\ \nu})$, where $G^\sigma_{\ \mu\nu}$ is the affine connection determined by $g^{\mu\nu}$. Finally we use the simplest choice of the tetrads, i.e. $e^{a}_{\ \mu}=\delta^a_{\ \mu}+  \delta^a_{\ i}\delta^0_{\ \mu} \, v_i$ and $e_{a}^{\ \mu}=\delta_a^{\ \mu} -  \delta_a^{\ 0}\delta_i^{\ \mu} \, v_i$.

We next consider the limit of very slow rotation, i.e. with $\omega$ being small
and expand the Lagrangian up to the order of $\hat{O}(\omega)$. After some lengthy but straightforward calculations, one arrives at the following result:
\begin{eqnarray}
\mathcal{L}=\psi^\dagger \left[ i\partial_0+i\gamma^0\vec{\gamma}\cdot\vec{\partial} + (\vec{\omega}\times\vec{x})\cdot(-i \vec{\partial})+\vec{\omega}\cdot \vec{S}_{4\times 4} \right ]\psi \,\,
\end{eqnarray}
where $\vec{S}_{4\times 4}=\frac{1}{2}  \left (
\begin{array}{cc}
\vec{\sigma} & 0 \\
   0 & \vec{\sigma}
\end{array}      \right )$ is the spin operator with $\vec{\sigma}$ the usual Pauli matrices. We note the last two terms in the above bracket may be interpreted as effective  polarization term $\vec \omega \cdot \vec J$, with total angular momentum $\vec J$ consisting of an orbital term and a spin term. The rotational velocity $\vec \omega$ serves as an effective ``chemical potential'' for total angular momentum of the system.

The next step is to find the ``natural'' eigenstates in this rotating frame, in parallel to the usual plane-wave spinor eigenstates in normal frame~\cite{Vilenkin:1980zv,Ambrus:2014uqa,rotating}. We first write the corresponding Hamiltonian in momentum space:
\begin{eqnarray} \label{eq_H}
\hat{H} =\gamma^0 (\vec{\gamma}\cdot \vec{p}+m )  -\vec{\omega}\cdot(\vec{x}\times \vec{p} +   \vec{S}_{4\times 4}  )
 =\hat{H}_0-\vec{\omega}\cdot\hat{\vec{J}}   \,\,
\end{eqnarray}
We use the cylindrical spatial coordinates $(r,\theta,z)$ with $\vec{\omega}=\omega \hat{z}$ and with $r,\theta$ being transverse radial position and azimuthal angle. It can be easily checked that the complete set of commutating operators consists of $\hat{H}$, $\hat{p}_z$, $\hat{\vec{p}}_{\, t}^{\, 2}$, $\hat{J}_z$, and $\hat{h}_t\equiv \gamma^5 \gamma^3
\vec{p}_t\cdot \vec{S}_{4\times 4}$\cite{baha}. The last one is a sort of reduced helicity operator on transverse plane.  One can therefore label the eigenstates of the above Hamiltonian by a set of corresponding eigenvalues: energy $E$, z-momentum $k_z$, transverse momentum magnitude $k_t$, z-angular-momentum quantum number $n=0,\pm1,...$, and ``transverse helicity''  $s=\pm$. The four solutions of spinor eigenstates are given by the following:
\begin{eqnarray}
 u_{k_z,k_t,n,s} = \sqrt{\frac{E_k+m}{4E_k}} e^{i k_z z} e^{i n \theta}
\left(
\begin{array}{cccc}
J_n(k_t r)\\
s\, e^{i\theta}J_{n+1}(k_t r) \\
\frac{k_z-i s\, k_t}{E_k+m}J_n(k_t r) \\
\frac{-s\, k_z+ik_t}{E_k+m}e^{i\theta}J_{n+1}(k_t r)\\
\end{array}
\right)  \,
\end{eqnarray}
\begin{eqnarray}
 v_{k_z,k_t,n,s} = \sqrt{\frac{E_k+m}{4E_k}} e^{-i k_z z} e^{i n \theta}
\left(
\begin{array}{cccc}
\frac{k_z-i s\, k_t}{E_k+m}J_n(k_t r) \\
\frac{s\, k_z-ik_t}{E_k+m}e^{i\theta}J_{n+1}(k_t r)\\
J_n(k_t r)\\
- s\, e^{i\theta}J_{n+1}(k_t r) \\
\end{array}
\right)  \,
\end{eqnarray}
where $E_k\equiv \sqrt{k_z^2+k_t^2+m^2}$ and $J_n(x)$ are $n$-th Bessel functions of the first kind. The energy eigenvalues
are simply $E=\pm E_k -(n+1/2)\omega$ with the plus (minus) for $u$ ($v$) spinor states respectively. The last term, i.e. $-(n+1/2)\omega$, is the ``rotational polarization energy''.  Clearly these results are the counterpart in rotating frame of the usual plane wave spinor states in   non-rotating frame. With these states as basis one can then compute various quantitates of interest using the standard thermal field theory method.

Finally we introduce an effective interaction  that takes the generic form of four-fermion contact vertex:
\begin{eqnarray}
{\cal L}_{I_{eff}} = G(\bar\psi \psi)^2+G_d(i\psi^T C\gamma^5\psi)(i\psi^\dagger C\gamma^5\psi^*)
\end{eqnarray}
The first term is a fermion-anti-fermion scalar-channel coupling while the second term is a di-fermion  scalar-channel coupling, with $G$ and $G_d$ the corresponding coupling constants. The above relativistic form of effective interaction is  the Nambu-Jona-Lasinio (NJL) model. It shall be  emphasized that essentially the same physics is applicable   to many other fermion systems (such as pairing in cold fermionic gases and conventional superconductor, etc). For specific application to chiral condensation and color superconductivity in QCD matter, the pertinent color/flavor indices and structures can be easily added to the above interaction (see e.g.~\cite{Klevansky}).

{\it Chiral Condensation in Rotating Matter.---} Let us first consider the chiral condensation which is a fermion-anti-fermion pairing phenomenon. Note for this pairing state, the spatial angular momentum (for the relative orbital motion) $L=1$ while the spin $S=1$, with the total angular momentum $J=0$ for the   fermion-anti-fermion pair. Following the standard mean-field method, one introduces the expectation value  $\left<\bar\psi \psi \right>$ that gives rise to a mean-field mass gap $M=m - 2G \left<\bar\psi \psi \right>$. Note that due to rotation, the system is no longer homogeneous and  the $M$ as well as $\left<\bar\psi \psi \right>$ become dependent on spatial coordinate --- specifically dependent only on $r$ by virtue of symmetry. Using the mean-field propagator one can compute the grand potential of the system:
\begin{eqnarray}
\Omega &=& \int d^3\vec r {\bigg \{}  \frac{(M-m)^2}{4G}  -\frac{N_f N_c}{16\pi^2}\sum_{n} \int d k_t^2 \int dk_z \nonumber \\
 && \times \quad [J_n(k_t r)^2+J_n(k_t r)^2] \nonumber \\
 && \times T {\bigg [} \ln\left(1+e^{(\epsilon_n-\mu)/T}\right)+\ln\left( 1+e^{-(\epsilon_n-\mu)/T}\right)  \nonumber  \\
&& \,\, + \ln\left(1+e^{(\epsilon_n+\mu)/T}\right)+\ln\left( 1+e^{-(\epsilon_n+\mu)/T}\right)
 {\bigg ]}\ {\bigg \}} \qquad
\end{eqnarray}
In the above the mean-field quasiparticle dispersion $\epsilon_n$ is given by $\epsilon_n=\sqrt{k_z^2+k_t^2+M^2}-(n+\frac{1}{2})\omega$.
The mean-field chiral condensate (or equivalently the mass gap $M$) at given values of temperature $T$, chemical potential $\mu$ and rotation $\omega$, can then be determined from the usual gap equation through variation of the order parameter: $\frac{\delta \Omega}{\delta M(r)}=0$ and $\frac{\delta^2 \Omega}{\delta M(r)^2}>0$.  We will numerically solve the gap equation for the case of $N_f=2$ and $N_c=3$ and present the results below.
For the parameters $G$, $G_d$ and a cutoff scale $\Lambda$ of this model, we  choose the standard values  (see e.g. \cite{Klevansky}).

\begin{figure}[!hbt]
\begin{center}
\includegraphics[scale=0.45]{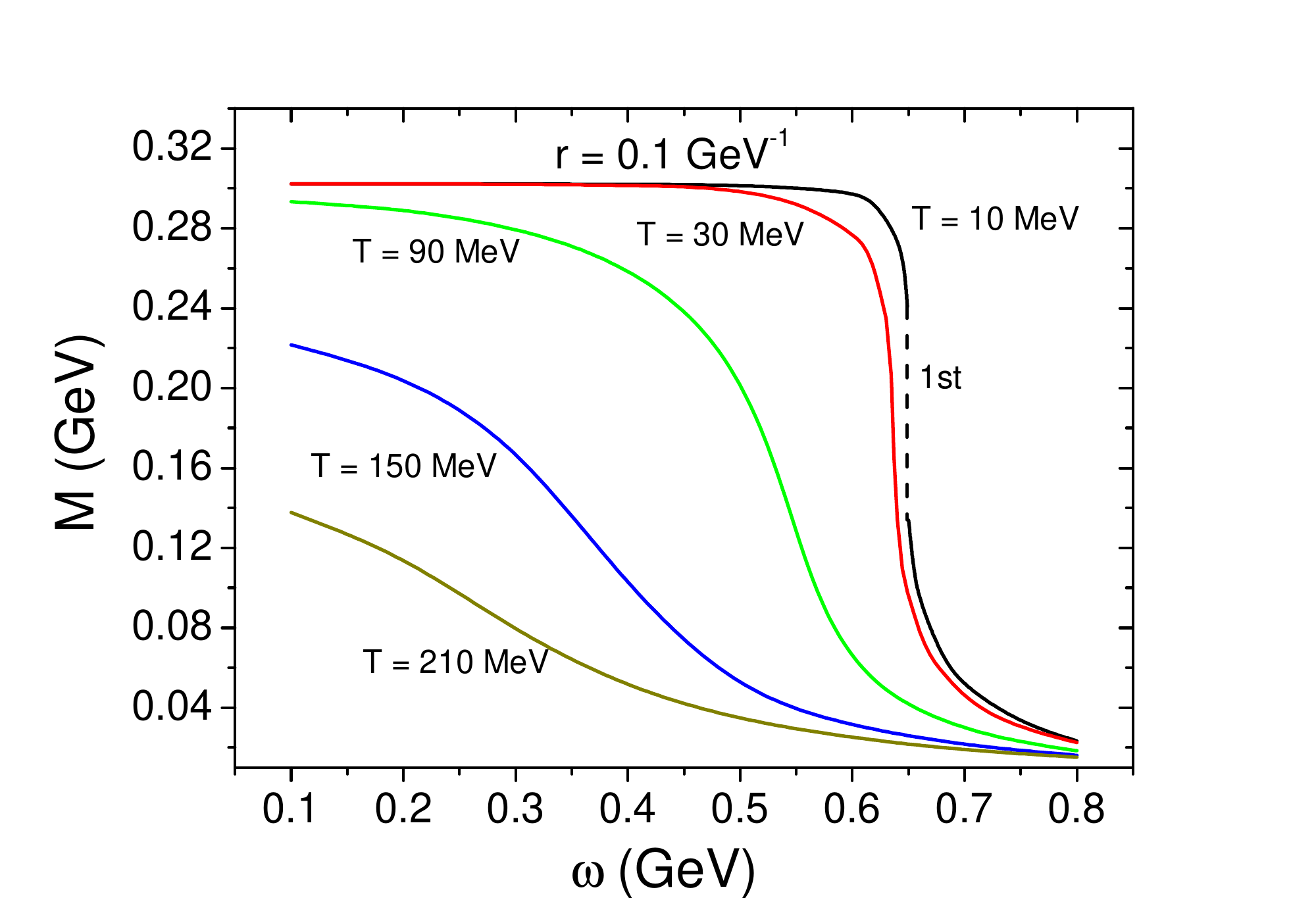}
\caption{The mean-field mass gap $M$ (at radius $r=0.1\rm GeV^{-1}$) as a function of $\omega$ for various fixed value of $T$.}
\vspace{-0.5cm}
\label{fig_omega}
\end{center}
\end{figure}

\begin{figure}[!hbt]
\begin{center}
\includegraphics[scale=0.45]{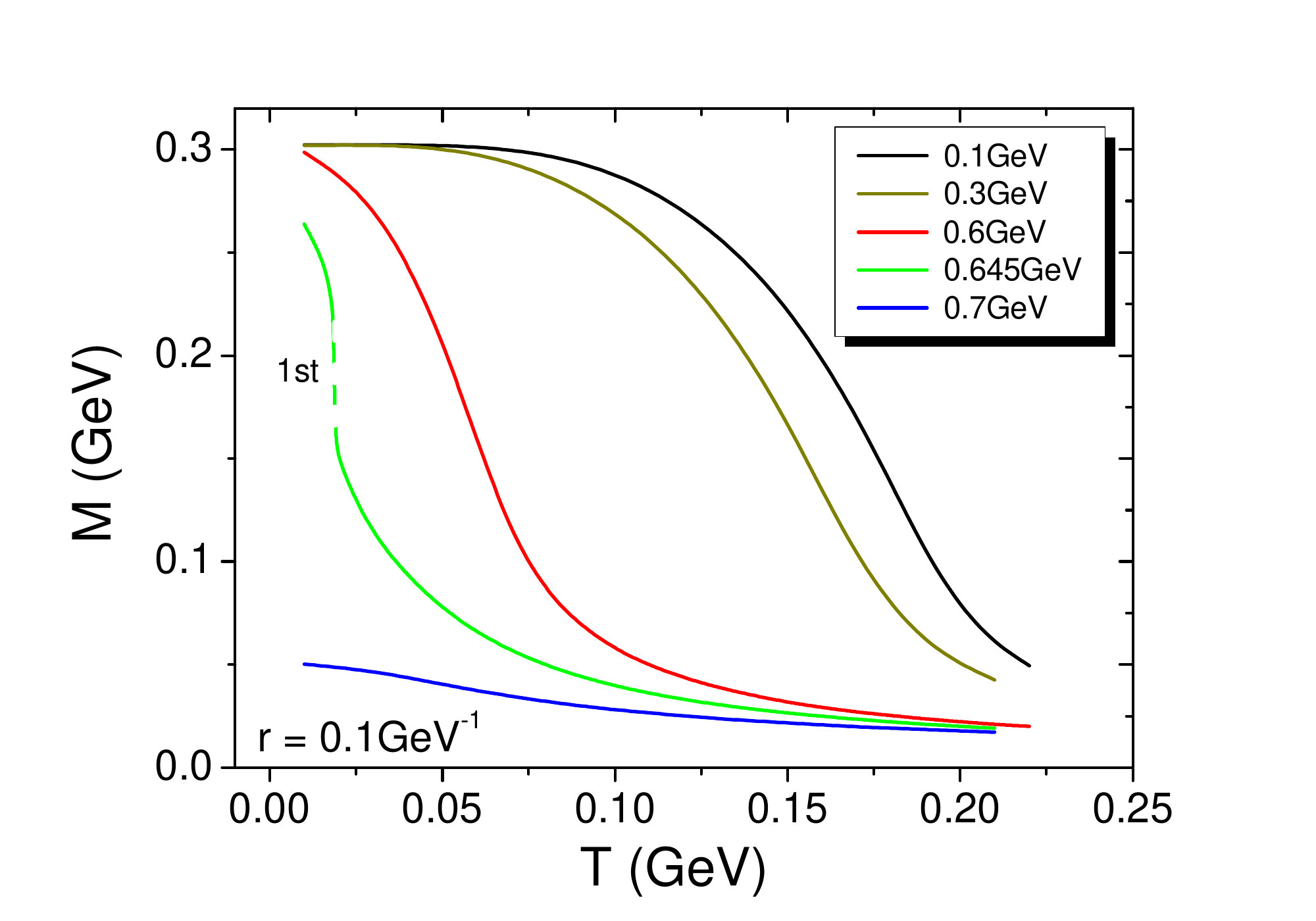}
\caption{ The mean-field mass gap $M$ (at radius $r=0.1\rm GeV^{-1}$) as a function of $T$ for various fixed value of $\omega$.}
\vspace{-0.75cm}
\label{fig_t}
\end{center}
\end{figure}

Let us focus on the zero density case (i.e. $\mu=0$) and study how the mass gap changes with $T$ and $\omega$. As already pointed out, the condensate will depend on the transverse radius $r$: we have found that the mass gap $M$ smoothly decreases with $r$ . In the following we will show results for a particular value of $r$ for simplicity.
 In Fig.~\ref{fig_omega} we show $M$ (at radius $r=0.1\rm GeV^{-1}$) as a function of $\omega$ for various fixed value of  $T$. At all values of temperature, the mass gap decreases with increasing values of $\omega$: this clearly confirms the rotational suppression effect on the quark-anti-quark pairing in the chiral condensate. We also see that at low temperature the chiral condensate experiences a first-order transition when $\omega$ exceeds a critical value $\omega_c$, while at high temperature the chiral condensate vanishes with increasing $\omega$ via a smooth crossover. The $\omega_c$ decreases with increasing temperature.
 In Fig.~\ref{fig_t} we show $M$ (at radius $r=0.1\rm GeV^{-1}$) as a function of $T$ for various fixed value of $\omega$. At very small $\omega$, the mass gap decreases smoothly toward zero with increasing temperature, indicating a smooth crossover transition as expected. However when $\omega$ becomes large, the transition becomes stronger and stronger, eventually becoming a first-order transition as signaled by a sudden jump. The transition temperature $T_c$ becomes smaller at larger $\omega$.  These results could be understood by considering $\omega$ as a sort of ``chemical potential'' for angular momentum. Indeed this is evident from Eq.(\ref{eq_H}): the term $\vec \omega \cdot \hat{\vec J}$ is in direct analogy to a term $\mu\cdot \hat{Q}$ for a conserved charge $\hat{Q}$. It is therefore not surprising that the phase transition behavior at finite $\omega$ is very similar to that at finite $\mu$ in the same model.

\begin{figure}[!hbt]
\begin{center}
\includegraphics[scale=0.45]{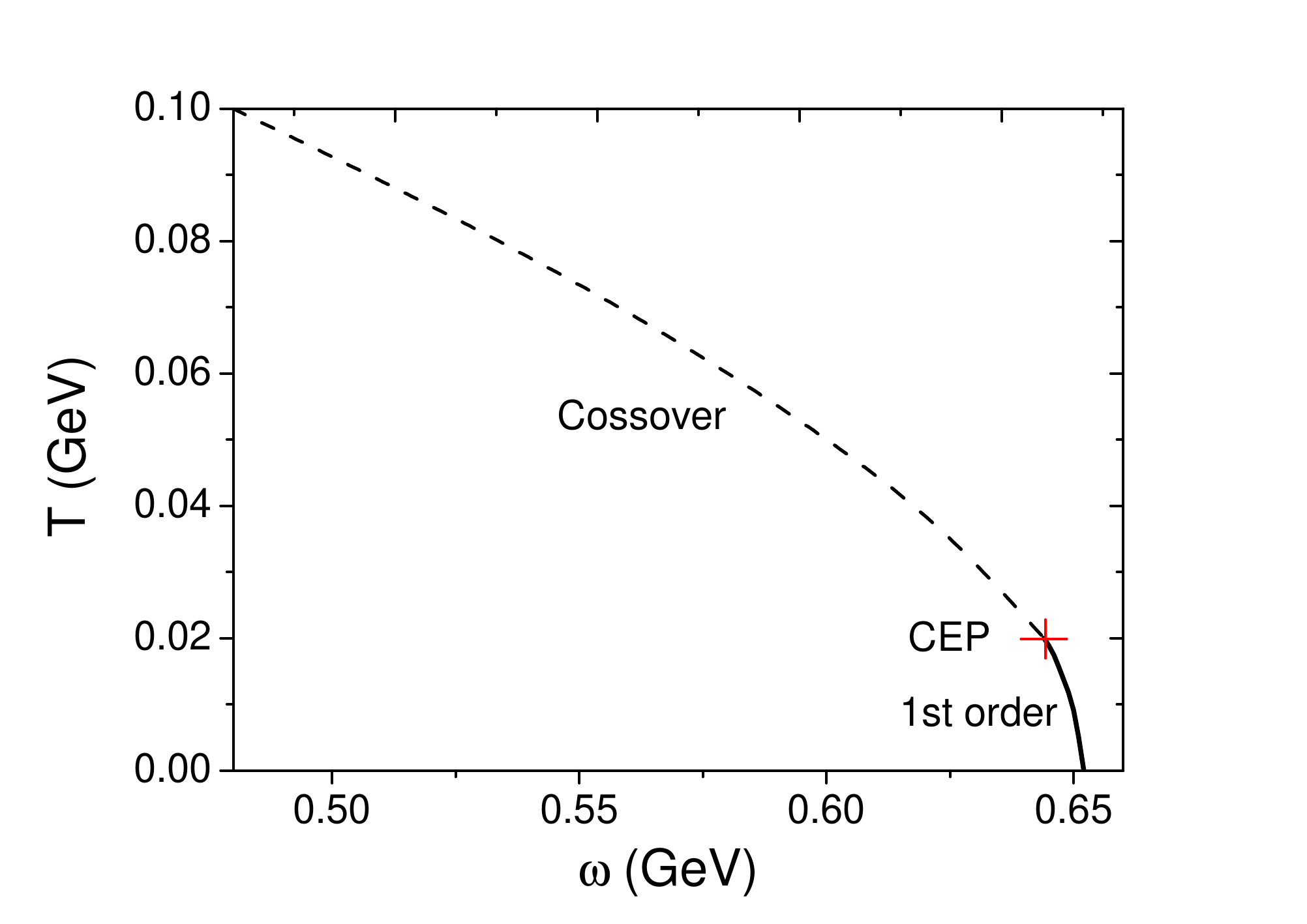}
\caption{The phase diagram on $T$-$\omega$ plane (see text).}
\vspace{-0.75cm}
\label{fig_phase}
\end{center}
\end{figure}

With the above observation, it is tempting to envision a new phase diagram of the chiral phase transition on the $T-\omega$ parameter space: see Fig.~\ref{fig_phase} (as computed from the present model). It features a chiral-symmetry-broken phase at low temperature and slow rotation while a chiral-symmetry-restored phase at high temperature and/or rapid rotation. A smooth crossover  transition region at high $T$ and low $T$ and a first-order transition line at low $T$ and high $\omega$ are connected by a new {\em critical end point}. Given the present model parameters, this critical point is located at $T_{CEP}=0.020\rm GeV$ and $\omega_{CEP}=0.644\rm GeV$. As already discussed previously, the ``rotational suppression'' of the scalar condensate is a quite generic effect. It is conceivable that    similar phase transition behaviors under rotation would also occur in other dynamical models for studying chiral condensate.

\begin{figure}[!hbt]
\begin{center}
\includegraphics[scale=0.45]{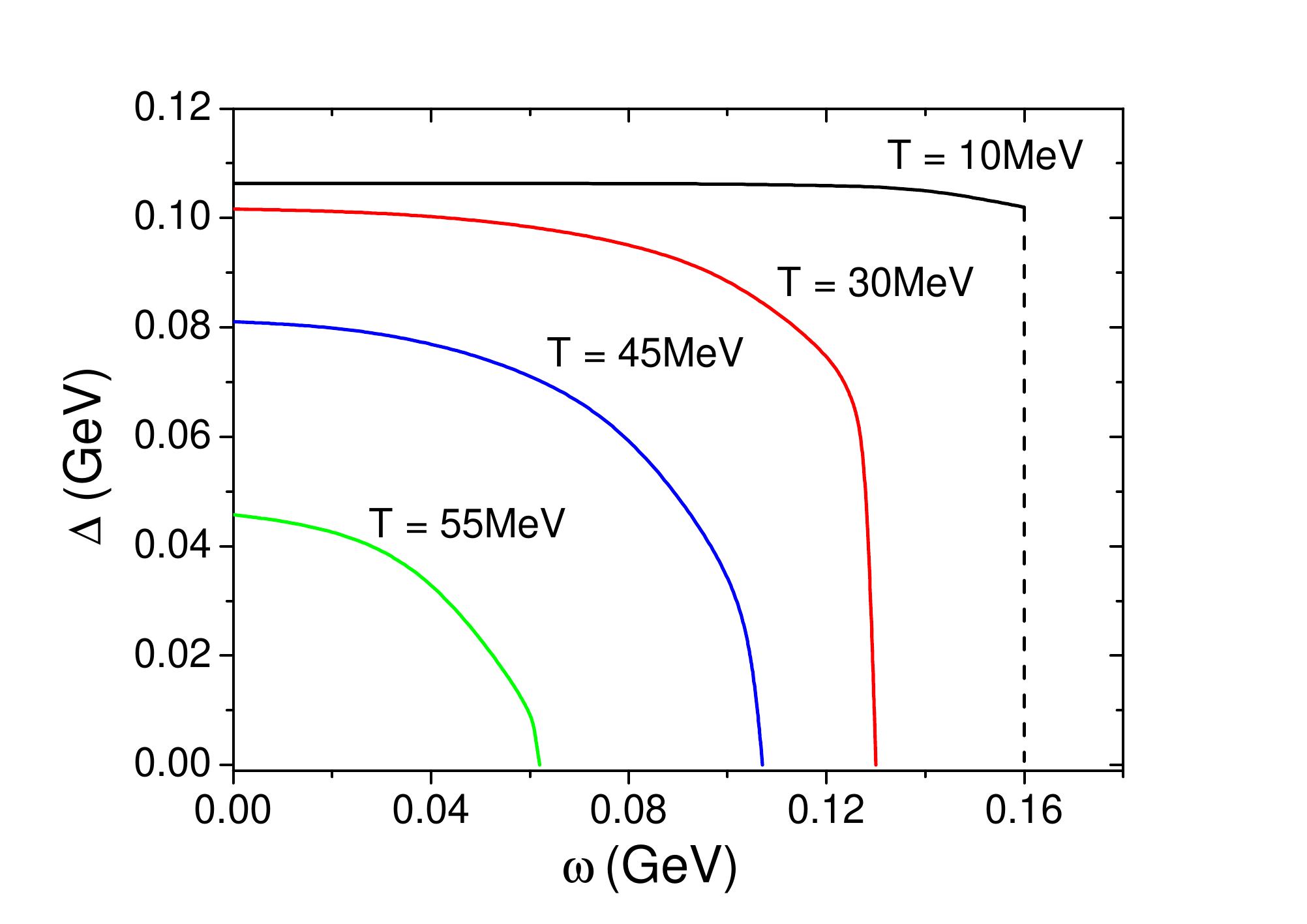}
\caption{The mean-field diquark condensate  $\Delta$ (at radius $r=0.1 \rm GeV^{-1}$) as a function of $\omega$ for  several values of $T$ and fixed value of $\mu=400\rm MeV$.}
\vspace{-0.5cm}
\label{fig_diquark}
\end{center}
\end{figure}

{\it  Superconducting Pairing in Rotating Matter.---} To demonstrate that the ``rotational suppression'' of the scalar condensate is a generic effect, we also study another quite different type of pairing: the fermion-fermion (rather than the fermion-anti-fermion) superconducting pairing phenomenon in the presence of rotation. In the QCD context, this is the color superconductivity at high density and low temperature (see e.g. \cite{Alford} for a recent review).  Quite different from the chiral condensate, the diquark pairing   state has   the spatial angular momentum (for the relative orbital motion) $L=0$ while the total spin $S=0$ (i.e. antisymmetric combination of the two individual quark spins), again with the total angular momentum $J=0$ for the pair.  We use the same NJL model and for simplicity we focus on the low-temperature high-density region where the chiral symmetry is already restored. Assuming a mean-field 2SC diquark condensate
$\Delta \epsilon^{\alpha\beta 3}\epsilon_{ij}=-2G_d\left< i  \psi^\alpha_i C \gamma^5 \psi^\beta_j \right> $   the grand potential in this case is given by:
\begin{eqnarray}
\Omega &=& \int d^3\vec r {\bigg \{}   \frac{\Delta^2}{4G_d}  -\frac{1}{16\pi^2}\sum_{n} \int d k_t^2 \int dk_z \nonumber \\
 && \times \quad [J_n(k_t r)^2+J_n(k_t r)^2] \nonumber \\
 && \times N_f T {\bigg [}
 (N_c-2)\left(\ln\left(1+e^{\epsilon_n^{+}/T}\right)+\ln\left( 1+e^{-\epsilon_n^{+}/T}\right)\right.  \nonumber  \\
&& \,\, \left.+ \ln\left(1+e^{\epsilon_n^{-}/T}\right)+\ln\left( 1+e^{-\epsilon_n^{-}/T}\right)\right) \nonumber \\
&& \,\, + 2 \left(\ln\left(1+e^{\epsilon_n^{\Delta+}/T}\right)+\ln\left( 1+e^{-\epsilon_n^{\Delta+}/T}\right)\right.  \nonumber  \\
&& \,\, \left.+ \ln\left(1+e^{\epsilon_n^{\Delta-}/T}\right)+\ln\left( 1+e^{-\epsilon_n^{\Delta-}/T}\right)\right)
 {\bigg ]}\ {\bigg \}} \qquad
\end{eqnarray}
In the above the mean-field quasiparticle dispersion $\epsilon_n^\pm$ and $\epsilon_n^{\Delta\pm}$ is given by $\epsilon_n^{\pm}=(\sqrt{k_z^2+k_t^2+m^2}\pm \mu)-(n+\frac{1}{2})\omega$ and
$\epsilon_n^{\Delta\pm}=[(\sqrt{k_z^2+k_t^2+m^2}\pm \mu)^2+\Delta^2]^{\frac{1}{2}}-(n+\frac{1}{2})\omega$.\\
The mean-field diquark condensate   $\Delta$ at given values of temperature $T$, chemical potential $\mu$ and rotation $\omega$, can then be determined from the self-consistency equation through variation of the order parameter: $\frac{\delta \Omega}{\delta \Delta(r)}=0$ and $\frac{\delta^2 \Omega}{\delta \Delta(r)^2}>0$.  By numerically solving the equation, we show in Fig.~\ref{fig_diquark} the   $\Delta$ (at radius $r=0.1 \rm GeV^{-1}$) as a function of $\omega$ for several values of $T$ and fixed $\mu=400\rm MeV$.  One can see that with increasing $\omega$, the diquark condensate always decreases  toward zero, through a 1st-order transition at low $T$ while a smooth crossover at higher $T$.  This result again confirms the generic rotational suppression effect on the scalar diquark  pairing.

{\it Summary and Discussions.---} In summary, we have found a generic rotational suppression effect on the fermion pairing state with zero angular momentum. This effect is demonstrated for two well-known pairing phenomena in QCD matter, namely the chiral condensate and the color superconductivity. The scalar pairing states in these two examples, while different in many aspects, are both found to be reduced with increasing rotation of the system. In the case of chiral phase transition, we have identified the phase boundary with a critical point on the $T-\omega$ parameter space.

The rotational effects on pairing phase transitions may bear interesting implications for a number of physics systems. The phase diagram of QCD matter on $T-\omega$ plane could be quantitatively explored by ab initio lattice simulations which has recently become feasible~\cite{Yamamoto:2013zwa}.  In heavy ion collisions there is sizable global angular momentum carried by the hot dense matter (as recently computed in e.g. \cite{Jiang:2016woz}): such rotational motion may cause the chiral restoration to occur at lower temperature as our results imply, and may bear measurable consequences (e.g. for dilepton emissions).  In the case of neutron stars, the dense QCD matter is under global rotation which may reduce the chiral as well as diquark or nucleon-nucleon pairings and may affect the moment of inertia for such stars~\cite{Berti, Demorest}. In the non-relativistic domain,  the cold fermionic gas is an ideal place to study the rotational suppression effect on the fermion pairing and the very interesting BCS-BEC crossover phenomenon~\cite{stringari, hui, urban, iskin}. Finally, while in this paper we limit ourselves to the study of slow rotation effects, it is worth commenting that highly nontrivial pairing phases (other than the scalar condensate considered in the present study) may arise in a very rapidly rotating system. Rapid global rotation will generally favor pairing states with {\em nonzero} angular momentum, and one could imagine the emergence of phases with such higher spin condensate. For example, in dense quark matter, spin-1 diquark condensate may become more favorable than the scalar diquark condensate  when $\omega$ becomes larger than certain value. There is also the possibility of inhomogeneous phase where condensate forms vortices carrying collective angular momentum (in analogy to the superconductor under magnetic fields). If a system has elementary excitations with nonzero spins (e.g. vector mesons in QCD system), then one may imagine the possibility that such excitations will have their masses reduced with increasing rotation and may become massless thus causing instability with strong enough rotation.
These are all interesting problems to be investigated in the future.

\vspace{0.2in}

 {\bf Acknowledgments.}
The authors thank K. Fukushima, X.-G. Huang, D. Kharzeev, L. McLerran, M. Stephanov, H.-U. Yee, and P. Zhuang for  discussions.  This material is based upon work supported by the U.S. Department of Energy, Office of Science, Office of Nuclear Physics, within the framework of the Beam Energy Scan Theory (BEST) Topical Collaboration. The work is also supported in part by the National Science Foundation under Grant No. PHY-1352368.
JL is
grateful to the RIKEN BNL Research Center for partial support.

\end{document}